\documentclass[a4paper,12pt]{article}
\usepackage{amsthm,amssymb,amsmath,url,amsfonts,mathrsfs,bm}
\theoremstyle{definition}

\newcommand\apc{\mathrel{%
  \ooalign{\raise0.0 ex\hbox{$\subset$}\cr\hidewidth\raise
0.2ex\hbox{\scalebox{0.7}{$\scriptstyle\app$}}\hidewidth\cr}}}
\newcommand\sic{\mathrel{%
  \ooalign{\raise0.0 ex\hbox{$\subset$}\cr\hidewidth\raise
0.2ex\hbox{\scalebox{0.7}{$\scriptstyle\sim$}}\hidewidth\cr}}}

\renewcommand{\AA}{{\cal A}}

\newcommand{\bea}{\begin{eqnarray*}}
\renewcommand{\v}{\mathbf{v}}
\renewcommand{\u}{\mathbf{u}}
\renewcommand{\k}{\mathbf{k}}

\newcommand{\hx}{\hat x}
\newcommand{\Rtn}{\mathbb{R}^{3N}}
\newcommand{\hv}{\hat v}
\newcommand{\eea}{\end{eqnarray*}}
\newcommand{\sa}{$\sigma$-algebra }

\newcommand{\dg}{\dagger}
\newcommand{\A}{\mathbb{A}}

\newcommand{\Pe}{Poincar\'e }

\newcommand{\V}{\mathbb{V}}
\newcommand{\B}{\mathbb{B}}

\newcommand{\td}{\tilde}

\newcommand{\app}{\approx}

\newcommand{\Ra}{\Rightarrow}

\newcommand{\XX}{\mathcal{X}}

\newcommand{\ga}{\gamma}

\newcommand{\eal}{\end{align*}}

\renewcommand{\ss}{\subset}

\newcommand{\tit}{\textit}
\newcommand{\Lam}{\Lambda}

\newcommand{\ity}{\infty}

\newcommand{\mo}{{-1}}

\newcommand{\bg}{\begin}

\newcommand{\en}{\end}

\newcommand{\tx}{\text}

\newcommand{\na}{\nabla}
\newcommand{\De}{\Delta}

\newcommand{\sm}{\setminus}

\newcommand{\HH}{{\cal H}}

\newcommand{\VV}{\mathcal{V}}

\newcommand{\om}{\omega}

\newcommand{\se}{\subseteq}

\renewcommand{\j}{\mathbf{j}}

\newcommand{\C}{\mathbb{C}}

\newcommand{\R}{\mathbb{R}}

\newcommand{\FF}{{\cal F}}
\newcommand{\Om}{\Omega}

\newcommand{\si}{\sigma}

\newcommand{\PP}{{\cal P}}

\newcommand{\X}{\mathbb{X}}

\newcommand{\K}{\mathbb{K}}

\newcommand{\la}{\langle}
\newcommand{\de}{\delta}
\newcommand{\ra}{\rangle}

\newcommand{\n}{{\bf n}}

\newcommand{\p}{{\bf p}}

\newcommand{\ba}{{\bm \alpha}}

\usepackage{graphicx}
\begin{document}
\begin{large}
\title{\bf {Asymptotic velocities in quantum and Bohmian mechanics}}
\end{large}
\author{Bruno Galvan\footnote{e-mail: b.galvan@virgilio.it}\\ \small via Melta 16, 38121 Trento, Italy.}
\maketitle
\begin{abstract}
In this paper the relations between the asymptotic velocity operators of a quantum system and the asymptotic velocities of the associated Bohmian trajectories are studied. In particular it is proved that, under suitable conditions of asymptotic regularity, the probability distribution of the asymptotic velocities of the Bohmian trajectories is equal to the one derived from the asymptotic velocity operators of the associated quantum system. It is also shown that in the relativistic case the distribution of the asymptotic velocities of the Bohmian trajectories is covariant, or equivalently, it does not depend on a preferred foliation (it is well known that this is not the case for the structure of the Bohmian trajectories or for their spatial distribution at a finite time). This result allows us to develop a covariant formulation of relativistic Bohmian mechanics; such a formulation is proposed here merely as a mathematical possibility, while its empirical adequacy will be discussed elsewhere.

\end{abstract}

\section{Introduction}

Bohmian mechanics \cite{bohm,hol,dbook,dbook2,qe} allows us to derive a set of trajectories in configuration space from the wave function of a quantum system of particles, and this set is naturally endowed with a probability measure. These trajectories and the resulting probability space will be referred to as the \tit{Bohmian trajectories} and the \tit{Bohmian space}, respectively.

The (forward) asymptotic velocity operators of a system of quantum particles are defined by the limit
\begin{equation} \label{l1} 
\lim_{t \to \ity}\frac {\hat U^\mo(t) \hat x \hat U(t)}{t},
\end{equation}
where $\hat x$ is the vector of the position operators of the particles and $\hat U(t)$ is the time evolution operator. The asymptotic velocity of a (Bohmian) trajectory $k(t)$ is the limit
\begin{equation} \label{l2} 
\lim_{t \to \ity} \frac{k(t)}{t}.
\end{equation}
In this paper, a quantum system will be said to be \tit{asymptotically regular} if the limit (\ref{l1}) exists and, in the case of a relativistic system, if moreover the asymptotic velocities operators are covariant. Analogously, a Bohmian space is said to be \tit{asymptotically regular} if the limit (\ref{l2}) exists for almost all the Bohmian trajectories. The asymptotic regularity of quantum systems and Bohmian spaces has been proved in various situations (see section \ref{se4}).

By assuming the asymptotic regularity of a quantum system and of the related Bohmian space, the following two results are proved in this paper: (a) the probability distribution of the asymptotic velocities of the Bohmian trajectories is equal to the probability distribution derived from the asymptotic velocity operators of the associated quantum system, and (b) for a relativistic system the probability distribution of the asymptotic velocities of the Bohmian trajectories is covariant.

The result (b) is remarkable because it is well known that in general a relativistic Bohmian space is not covariant \cite{bohm}. This fact is usually expressed in coordinate-free terms by saying that constructing a relativistic Bohmian space requires selecting a preferred foliation of space-time \cite{hyper}. The result (b) states that, on the contrary, the distribution of the asymptotic velocities of the Bohmian trajectories does not depend on any preferred foliation. This result allows us to build a covariant Bohmian space, and therefore to formulate Bohmian mechanics in a covariant way. Such a formulation is proposed here merely as a mathematical possibility, and its empirical adequacy will be discussed in a future paper.

In the literature, the asymptotic behavior of the Bohmian trajectories has been studied into the details in \cite{asym} for a single non-relativistic particle. The results of that paper, which are consistent with the above assumptions and results, are recalled in section \ref{se4}. 

The plan of the paper is the following. In Section \ref{se2} the definition of trajectory space and some related notions are presented, where a trajectory space is considered here as a generic set of trajectories in configuration space endowed with a probability measure. In Section \ref{se3} some notions relative to quantum systems and asymptotic velocity operators are presented. In Section \ref{se4} it is shown how to derive the Bohmian space from a quantum system, and the two announced results (a) and (b) are proved. In Section \ref{se5} the covariant Bohmian space is presented.

\section{Trajectory spaces} \label{se2}

Let $\X:=\R^{3N}$ be the configuration space of a $N$-particles system, with Borel \sa $\XX$. Let moreover $\K:=C^0(\R, \X)$ be the set of the continuous trajectories from $\R$ to $\X$; a trajectory of $\K$ can also be considered as an $N$-tuple of trajectories in $\R^3$. Let $\pi_t: \K \ni k \mapsto k(t) \in \X$ be the configuration map; any subset $\B \se \K$ is naturally endowed with the \sa $\si(\B):=\si(\{\pi_t|_\B\}_{t \in \R})$, which is the Borel \sa generated by the topology of the uniform convergence on compact sets \cite{bau}. In the appendix (proposition \ref{prop2}) it is proved that for two generic sets $\A, \B \se \K$ with $\B \se \A$, we have
\bg{equation} \label{laste}
\si(\B)=\{\B \cap A: A \in \si(\A)\}.
\en{equation}
This implies in particular that if $\B \in \si(\A)$ then $\si(\B) \se \si(\A)$.

A \tit{trajectory space} is a probability space of the type
\begin{equation} \label{tras}
(\B, \si(\B), P),
\end{equation}
where $\B \se \K$.

Given a trajectory space $(\B, \si(\B), P)$, for every $t \in \R$ the map $\pi_t|_\B$ induces the probability measure
\begin{equation} \label{isd}
P_t:= P \circ \pi_t|_{\B}^\mo
\end{equation}
on $(\X, \XX)$, which is the instantaneous probability distribution of the particles on configuration space.

\tit{Asymptotic velocities.} Let $\V:=\Rtn$ be the space of velocities, with Borel \sa $\VV$. For $t > 0$ let us define the map
\begin{equation} \label{}
\eta_t:\K \ni k  \mapsto \frac{\pi_t(k)}{t} \in \V.
\end{equation}
Given a trajectory space $(\B, \si(\B), P)$ the map $\eta_t|_\B$ induces the probability measure
\bg{equation} \label{isc}
S_t := P \circ \eta_t|_\B^\mo
\en{equation}
on $(\V, \VV)$. Note that
\bg{equation} \label{ana}
S_t = P_t \circ h_t,
\en{equation}
where $h_t$ is the map $h_t:\V \ni v \mapsto t v \in \X$. In fact $\eta_t|_\B=h^\mo_t \circ \pi_t|_\B$, so that $S_t = P \circ \pi_t|_\B^\mo  \circ h_t = P_t \circ h_t$.

Let $\K_A$ be the subset of $\K$ composed of the trajectories $k$ for which
\begin{equation} \label{}
\lim_{t \to \ity} \eta_t(k) = :\eta_+(k)
\end{equation}
exists finite. The vector $\eta_+(k) \in \V$ is the \tit{asymptotic velocity} of the trajectory $k$. For any set $\B \se \K_A$ the map $\eta_t|_\B: \B \to \V$ is measurable for all $t > 0$, and converges to $\eta_+|_\B$ pointwise for $t \to \ity$; as a consequence also $\eta_+|_\B$ is measurable, that is 
\bg{equation} \label{send}
\si(\eta_+|_\B) \se \si(\B).
\en{equation}

Let $(\B, \si(\B), P)$ be a trajectory space, and let $\B_A:=\B \cap \K_A$. We say that the trajectory space is \tit{asymptotically regular} if $\B_A \in \si(\B)$ and $P(\B_A)=1$. In this case, due to equations (\ref{laste}) and (\ref{send}), we have $\si(\eta_+|_{\B_A}) \se \si(\B_A) \se \si(\B)$, and the map $\eta_+|_{\B_A}$ induces the probability measure
\begin{equation} \label{av}
S_+:= P \circ \eta_+|_{\B_A}^\mo
\end{equation}
on $(\V, \VV)$. This measure is the probability distribution of the asymptotic velocities, and will be referred to as the \tit{asymptotic measure} of the trajectory space. One can prove that \begin{equation} \label{prl1}
w-\lim_{t \to \ity} S_t=S_+,
\end{equation}
where the weak limit means that $\lim_{t \to \ity} \int_\V f dS_t = \int_\V f dS_+$ for all continuous bounded functions $f: \V \to \R$ \cite{billi}. In fact $f \circ \eta_t(k) \to f \circ \eta_+(k)$ for $P$-almost all $k \in \B$; moreover $|f \circ \eta_t (k)| \leq \sup_v |f(v)|$; according to the dominated convergence theorem we have $\int_\V f dS_t = \int_\X f \circ \eta_t  dP \to \int_\X f \circ \eta_+ dP = \int_\V f dS_+$.

While the asymptotic regularity of a trajectory space implies that the measure $S_t$ converges for $t \to \ity$, the opposite implication is not true. Consider for example the following trajectory space: $\X:=\R^3$, $\B:=\{k_\v(t):= R(\n, \om t)\v t: \|\v\| \in S^2 \}$, where $R(\n, \om t)$ is the 3-rotation of the angle $\om  t$ around the axis $\n \in S^2$, and $P$ is the measure induced by the normalized Lebesgue measure on $S^2$. Then $\eta_t(\B)=S^2$ and $S_t$ is the normalized Lebesgue measure concentrated on $S^2$. As a consequence $S_t$ is independent of $t$ and therefore converges for $t \to \ity$, but $\eta_t(k_\v) = R(\n, \om t)\v$ does not converge for any $\v$.

\tit{Relativistic trajectory spaces}. A \tit{world line} is a map $\k:\R \to \R^3$ satisfying the condition ($c=1$):
\bg{equation} \label{wl}
(t-s)^2 - [\k(t)- \k(s)]^2 \geq 0 \tx{ for all } t, s \in \R.
\en{equation}
Note that this condition implies that $\k$ is continuous. Let $\K_W$ be the subset of $\K$ composed of the trajectories $k=(\k_1, \ldots, \k_N)$ such that every $\k_i$ is a world line: we say that a trajectory space $(\B, \si(\B), P)$ is \tit{relativistic} if $\B \se \K_W$.

Let us study how a relativistic trajectory space transforms under the action of the \Pe group. Let $\PP$ denote the (proper orthocronous) \Pe group of transformations on $M:=\R \times \R^3$, and let the elements of $\PP$ be denoted by $g$. The graph of a world line is a subset of $M$, and its image under a \Pe transformation is still the graph of a world line. As a consequence the \Pe group acts naturally on the space of the world lines. Let us see explicitly how a world line transforms when $g$ is a boost of velocity $u<1$ along the $x$-axis (this is also preparatory for the study of the transformation of the asymptotic velocities). The graph of $\k$ is the set 
\begin{equation} \label{}
\bar \k := \{ (t, k_x(t), k_y(t), k_z(t)) \in M: t \in \R\}.
\end{equation}
The transformed graph is
\begin{equation} \label{trg}
g \bar \k := \{ (\ga [t-uk_x(t)], \ga[k_x(t)-ut], k_y(t), k_z(t)) \in M: t \in \R\},
\end{equation}
where $\ga=1/\sqrt{1-u^2}$. Let us define the function
\bg{equation}
\td s(t):= \ga [t-uk_x(t)].
\en{equation}
The function $\td s(t)$ is (i) continuous, (ii) increasing, and (iii) $\td s(t) \to \pm \ity$ for $t \to \pm \ity$. The continuity of $\td s(t)$ is obvious; it is increasing because
\bea
& & \td s(t_2)- \td s(t_1)  = \ga (t_2-t_1) - u [k_x(t_2) -k_x(t_1)] \geq \\
& & \geq \ga (t_2-t_1) - u \|\k(t_2) -\k(t_1)\| \geq (t_2-t_1) (\ga -u ) >0
\eea
for $t_1 < t_2$. From the above inequality one obtains
\[
\td s(t) > \td s(0) + t (\ga-u)  \tx{ for } t >0,
\]
from which the positive asymptotic limit can be deduced; the negative limit can be proved analogously. The function $\td s(t)$ is therefore invertible, and the inverse function $\td t(s)$ has the same three properties (i), (ii), and (iii) than $\td s(t)$. Due to these properties the graph (\ref{trg}) can equivalently be written as follows:
\[
g \bar \k = \{ (\ga [\td t(s)-uk_x(\td t(s))], \ga[k_x(\td t(s))-u \td t(s)], k_y(\td t(s)), k_z(\td t(s))) \in M: s \in \R\}.
\]
Since, by definition, 
\bg{equation} \label{feq1}
s=\ga [\td t(s)-uk_x(\td t(s))],
\en{equation}
one can also write
\begin{equation} \label{}
g \bar \k = \{ (s, \ga[k_x(\td t(s))-u\td t(s)], k_y(\td t(s)), k_z(\td t(s))) \in M: s \in \R\},
\end{equation}
from which one obtains that the transformed world line is
\begin{equation} \label{feq2}
g \k(s) = (\ga[k_x(\td t(s))-u\td t(s)], k_y(\td t(s)), k_z(\td t(s))).
\end{equation}

The action of $\PP$ is naturally extended to $\K_W$ by defining $[g(\k_1, \ldots, \k_N)](t):=(g\k_1(t), \ldots g\k_N(t))$. For $\B \se \K_W$ we define moreover $g \B:=\{ g k: k \in \B\}$ and $g \si(\B):=\{g K: K \in \si(\B)\}$. One can prove that 
\begin{equation} \label{}
g\si(\B)= \si(g \B).
\end{equation}
The sketch of proof is the following: one easily sees that the map $g^\mo:g \B \to \B$ is continuous in the topology of the uniform convergence on compact sets, and therefore it is $(\si(g\B), \si(\B))$-measurable. As a consequence $g K \in \si(g\B)$ for $K \in \si(\B)$, from which $g\si(\B) \se \si(g\B)$. Analogously one can prove that $g^\mo\si(g\B) \se \si(\B)$, which implies $\si(g\B) \se g \si(\B)$ and therefore $g\si(\B)=\si(g\B)$.

The measure $P$ transforms according to the usual push-forward of measures:
\begin{equation} \label{}
g P(K):= P(g^\mo K) \tx{ for all } K \in \si(g \B).
\end{equation}
The transformed trajectory space is therefore
\begin{equation} \label{}
g(\B, \si(\B), P):=(g \B, \si(g \B), gP).
\end{equation}

Let us consider now asymptotic velocities. Let $\K_{WA}:=\K_W \cap \K_A$. The range of $\eta_+|_{\K_{WA}}$ is the space $\V_1:=\{ \v \in \R^3: \|\v \| \leq 1\}^N$, with Borel \sa $\VV_1$. Let us prove this for $N=1$: for $\|\v\| \in \V_1$ the trajectory $\k(t) := \v t$ belongs to $\K_{WA}$, and $\eta_+(\k)=\v$, so that $\V_1 \se \eta_+(\K_{WA})$; on the other hand, for a generic world line $\k$ we have
\[
\left \| \lim_{t \to \ity} \k(t)/t \right \| = \lim_{t \to \ity} \|\k(t)\|/t = \lim_{t \to \ity} \|\k(t)- \k(0)\|/t \leq 1,
\]
and therefore $\eta_+(\K_{WA}) \se \V_1$. The case of a generic $N$ is a straightforward generalization. A consequence of the equality $\eta_+(\K_{WA}) = \V_1$ is that the support of the asymptotic measure of a relativistic trajectory space is contained in $\V_1$.

Let $\k$ be a world line such that $\lim_{t \to \ity}\k(t)/t=\v$. One can easily prove that $\lim_{t \to \ity}g\k(t)/t=\v$ if $g$ is a translation, and $\lim_{t \to \ity}g\k(t)/t=R\v$ if $g$ is a $3$-rotation $R$. Suppose that $g$ is a boost of velocity $u$ along the $x$-axis; from equations (\ref{feq2}) and (\ref{feq1}) one deduces that
\bea
& & \lim_{s \to \ity} g\k(s)/s= \left (\frac{\ga[k_x(\td t(s))-u\td t(s)]}{\ga [\td t(s)-uk_x(\td t(s))]}, \frac{k_y(\td t(s))}{\ga [\td t(s)-uk_x(t(s))]}, \frac{k_z(\td t(s))}{\ga [\td t(s)-uk_x(\td t(s))]}\right )= \\
& & = \left (\frac{v_x-u}{1 - u v_x}, \frac{v_y}{\ga (1 - u v_x)}, \frac{v_z}{\ga (1 - u v_x)} \right),
\eea
which is the usual transformation law for relativistic velocities under boosts. In general, the transformation law for the relativistic (and asymptotic) velocities under a Lorentz transformation $\Lam$ is the following: let $(u_0, \u)$ be a future-directed time-like or light-like four vector such that $\v=\u/u_0$; then $\Lam \v = \u'/u_0'$, where $(u'_0, \u')=\Lam(u_0, \u)$. A consequence of these results is that if $\k$ admits asymptotic velocity then $g\k$ admits asymptotic velocity as well, and for two world lines $\k$ and $\k'$ the equality $\lim_{t \to \ity}\k(t)/t=\lim_{t \to \ity}\k'(t)/t$ implies the equality $\lim_{t \to \ity}g\k(t)/t=\lim_{t \to \ity}g\k'(t)/t$.

The above results extend in a straightforward way to the trajectories of $\K_{WA}$, and again $k \in \K_{WA} \Ra g k \in \K_{WA}$  and $\eta_+(k) = \eta_+(k') \Ra \eta_+(gk) = \eta_+(gk')$. The action of the \Pe group can therefore be naturally defined on $\V_1$ as follows: if $v = \eta_+(k)$ we define
\begin{equation} \label{}
gv := \eta_+(gk).
\end{equation}

It is straightforward to prove that if $(\B, \si(\B), P)$ is asymptotically regular then $g(\B, \si(\B), P)$ is asymptotically regular as well. From the definition (\ref{av}) we deduce that the transformed asymptotic measure is
\begin{equation} \label{dav}
gS_+:=gP \circ \eta_+|_{g\B_A}^\mo.
\end{equation}
One can prove that:
\begin{equation} \label{dav2}
g S_+(V)=S_+(g^\mo V) \tx{ for all } V \in \VV_1.
\end{equation}
In fact
\[
gS_+(V)=gP(\eta^\mo_+(V) \cap g \B_A) = P(g^\mo \eta _+|_{\K_{WA}}^\mo (V) \cap  \B_A) = P(\eta^\mo_+(g^\mo V) \cap  \B_A) = S_+(g^\mo V).
\]
The equality (\ref{dav2}) holds true only if the trajectory space is asymptotically regular, in the following sense: suppose that the space is not asymptotically regular but nevertheless $S_t \to S_\ity$ for $t \to \ity$; in this case there is no guarantee that $gS_t \to gS_\ity$ for $t \to \ity$, where $gS_t :=g P_t \circ \eta_t|_{g\B_A}^\mo$ and $gS_\ity(V) := S_\ity(g^\mo V)$. See the supplementary material \cite{supp}.

\section{Quantum systems} \label{se3}

The considered quantum systems will be $N$-particle systems with Hilbert space $\HH$, Hamiltonian $\hat H$, and position operators $\hx=(\hat{\mathbf{x}}_1, \ldots, \hat{\mathbf{x}}_N)$. The Hamiltonian defines the time evolution operators $\hat U(t)=e^{-i\hat Ht}$ ($\hbar =1$), and the position operators define the PVM (projector valued measure) $\hat E$ on $(\X, \XX)$. If the system is relativistic, a representation $\hat U_g$ of the (covering group of the) \Pe group is also given, with the usual relation $\hat U^\mo_t = \hat U(t)$ between time translation and time evolution operators. A normalized vector $\psi \in \HH$ representing the state of the system at the time $t=0$ is singled out. Two examples of this type of quantum systems are given at the end of the section.

For $t > 0$ let us define the operators 
\begin{equation}
\hx_t:=\hat U^\mo(t) \hx \hat U(t) \tx{ and } \hat v_t := \frac{\hx_t}{t},
\end{equation}
and let $\hat E_t$ and $\hat F_t$ denote the PVMs on $(\X, \XX)$ defined by the operators $\hx_t$ and $\hat v_t$, respectively. One can prove that 
\bg{equation} \label{oop}
\hat F_t=\hat E_t \circ h_t.
\en{equation}
In fact for all $V \in \VV$ we have: $\hat F_t(V)=\chi_V(\hv_t)= \chi_V ( h_t^\mo (\hx_t))= \chi_V \circ h_t^\mo (\hx_t)= \chi_{h_t(V)} (\hx_t)= \hat E_t \circ h_t (V)$, where $\chi_V$ is the characteristic function of the set $V$.

Let us define moreover the following probability measures on $(\X, \XX)$:
\begin{equation} \label{prob}
R^\psi_t := \la \psi |\hat E_t( \cdot)| \psi \ra \tx{ and } Q^\psi_t:=\la \psi |\hat F_t( \cdot)|\psi\ra.
\end{equation}
From equation (\ref{oop}) it descends that 
\begin{equation} \label{}
Q^\psi_t = R^\psi_t \circ h_t,
\end{equation}

\tit{Asymptotic velocity operators}. We say that a (non-relativistic) quantum system is \tit{asymptotically regular} if the limit 
\bg{equation} \label{limit}
s-C_\ity- \lim_{t \to \ity} \hv_t =:\hv_+
\en{equation}
exists, and the operators $\hv_+$ are commuting self-adjoint operators. The limit (\ref{limit}) means that
\begin{equation} \label{l3}
s-\lim f(\hv_t)=f(\hv_+)
\end{equation}
for all the continuous functions $f:\V \to \R$ such that $f(v) \to 0$ for $\|v\| \to \ity$ (see \cite{dere}). The operators $\hv_+$ are the \tit{asymptotic velocity operators} of the system.

Let $\hat F_+$ denote the PVM on $(\V, \VV)$ defined by the operators $\hv_+$. If the system is relativistic, a further condition for the asymptotic regularity is required, namely that the support of $\hat F_+$ is contained in $\V_1$ and that $\hat F_+$ is covariant, that is:
\begin{equation} \label{cova}
\hat U_g^\mo \hat F_+(V) \hat U_g = \hat F_+ (g^\mo V) \tx{ for all } V \in \VV_1.
\end{equation}
This condition basically corresponds to the usual requirement of scattering theory that the ``in'' and ``out'' states transform as free states under the action of the \Pe group \cite{fong}.

If the quantum system is asymptotically regular the following probability measure can be defined on $(\V, \VV)$:
\bg{equation} \label{qap}
Q^\psi_+:=\la \psi |\hat F_+( \cdot)|\psi \ra.
\en{equation}
From equation (\ref{limit}) it descends that
\bg{equation} \label{l4}
 w-\lim_{t \to \ity} Q^\psi_t = Q^\psi_+.
\en{equation}
In fact one can prove that if the limit (\ref{l3}) holds true for all the continuous functions $f$ such that $f (v) \to 0$ for $\|v\| \to \ity$ then it holds true for all the continuous bounded function (see for example the proof of Theorem VIII.20 (b) in \cite{reed}). As a consequence
\[
\lim_{t \to \ity} \int f(v) dQ_t^\psi = \lim_{t \to \ity} \la \psi |\int f(v) dF_t^\psi| \psi \ra = \la \psi |\int f(v) dF_+^\psi| \psi \ra = \int f(v) dQ_+^\psi
\]
for all continuous bounded functions $f$, which proves equation (\ref{l4}).

Finally from condition (\ref{cova}) it descends trivially that the measure $Q^\psi_+$ is covariant, that is:
\begin{equation} \label{mcv}
Q^{U_g \psi}_+(V) = Q^\psi_+ (g^\mo V) \tx{ for all } V \in \VV_1.
\end{equation}

\tit{Examples.} For the non-relativistic case we consider a system of spin-0 particles with $\HH = L^2(\X)$,
\bg{equation}
\hat H = - \sum_{i=1}^N \frac{\De_i}{2 m_i} + V(\hat x),
\end{equation}
and the usual multiplication by the coordinates as position operators, so that $E(X)$ is the multiplication by the characteristic function of the set $X$. If $\hat V=0$ the asymptotic velocity operators are equal to the velocity operators $\hv:=(\hat{\mathbf{p}}_1/m_1, \ldots, \hat{\mathbf{p}}_N/m_N)$, where the $\hat{\mathbf{p}}_i$s are the momentum operators. For a single particle subjected to a potential admitting M\"oller operators $\hat \Om_\pm$ we have $\hv_+=\hat \Om_+ \hv \,\hat \Om_+^\dg$ \cite{dere}. More in general, it has been proved in \cite{dere} that the limit (\ref{limit}) exists for a wide class of potentials (including the Coulomb potential), and that $\hv_+$ is a vector of commuting self-adjoint operators which also commute with the Hamiltonian. 

For the relativistic case the example is a system of free fermions, with $\HH=L^2(\X) \times (\C^4)^{\otimes N}$, the free Dirac Hamiltonian as Hamiltonian, and the other generators of $\hat U_g$ as defined in \cite{taller}. The position operators are again the multiplication by the coordinates, which in this case are referred to as the \tit{standard} position operators \cite{taller}; in this case too $\hat E(X)$ is the multiplication by the characteristic function of the set $X$. The asymptotic velocity operators are equal to the classical velocity operators $\hv:=(\hat \p_1/\hat H_1, \ldots, \hat \p_N/\hat H_N)$, where $\hat H_i$ is the Dirac Hamiltonian of the $i$-th particle \cite{taller}. The covariance of $\hat F_+$ derives from the fact that the classical velocity operators transform as relativistic velocities, because any $(\hat H_i, \hat \p_i)$ is a four-vector; the details of the proof are omitted.

\section{Bohmian spaces} \label{se4}

A Bohmian space is a trajectory space generated by an associated quantum system in the following way. A probability density $\rho_t(x)$ and a probability current $j_t(x)=(\j_{1t}(x), \ldots, \j_{Nt}(x))$ must be derived from the wave function $\psi_t:=\hat U(t)\psi$, and they must satisfy the continuity equation
\bg{equation}
\dot \rho_t + \na \cdot j_t=0.
\end{equation}
For the two systems presented in the previous section the two densities are
\begin{equation} \label{}
\rho_t(x):=|\psi_t(x)|^2 \tx{ and } \j_{it}(x):=\frac{1}{m_i} \tx{Im} \,\psi_t^*(x) {\bm \na}_i \psi_t(x)
\end{equation}
for the non-relativistic system, and
\begin{equation} \label{}
\rho_t(x):=\psi^\dg_t(x) \psi_t(x) \tx{ and } \j_{i t}(x):=\psi^\dg _t(x) \hat \ba_i \psi_t(x)
\end{equation}
for the relativistic system. Note that in both the cases the measure $R^\psi_t$ defined by equation (\ref{prob}) is induced by the probability density $\rho_t$, that is $R^\psi_t(X) = \int_X \rho_t d^{3N}x$. The set $\B^\psi$ of the Bohmian space is composed of the solutions of the guiding equation
\bg{equation} \label{diff}
\dot k(t)=\frac{j_t(k(t))}{\rho_t(k(t))}.
\end{equation}
Since the guiding equation is a first order ODE, the map $\pi_t|_{\B^\psi}$ is invertible for all $t \in \R$ (see note\footnote{Actually it is possible that $\pi_t|_{\B^\psi}$ is not surjective because the guiding equation may have no solution for same initial configurations. However in \cite{global} it has been proved for the non-relativistic case that under suitable conditions for the wave function there is a solution of the guiding equation for $R^\psi_0$-almost all the initial configurations at the time $t=0$. For simplicity these technicalities are ignored here, and it is assumed that a solution of the guiding equation defined on the whole real axis exists for all the initial configurations.}); it will be assumed moreover without proof that the flux map $\pi_s \circ \pi_t|_{\B^\psi}^\mo: \X \to \X$ is measurable for all $s,t \in \R$. According to proposition \ref{prop1} in the appendix, this implies that $\pi_t|^\mo_{\B^\psi}$ is measurable for all $t \in \R$, and therefore the probability measure
\bg{equation} \label{mesp}
P^\psi:=R^\psi_t \circ \pi_t|_{\B^\psi}
\end{equation}
is well defined. It is well known that the flux generated by the Bohmian trajectories is equivariant, that is 
\begin{equation} \label{}
R^\psi_t \circ \pi_t|_{\B^\psi}= R^\psi_s \circ \pi_s|_{\B^\psi} \tx{ for all } t,s \in \R,
\end{equation}
and therefore the measure (\ref{mesp}) does not depend on  time. The Bohmian space associated with the wave function $\psi$ is therefore the trajectory space\footnote{For simplifying the notation in this section the \sa will be omitted.} 
\begin{equation} \label{}
(\B^\psi, P^\psi).
\end{equation}

The above definition of $P^\psi$ implies trivially that
\begin{equation} \label{eq1}
P^\psi_t = R^\psi_t \tx{ and } S^\psi_t = Q^\psi_t \tx{ for all } t \in \R,
\end{equation}
where $P^\psi_t$ and $S^\psi_t$ are the measures relative to the space $(\B^\psi, P^\psi)$ defined by the equations (\ref{isd}) and (\ref{isc}), respectively.

In the relativistic case one can prove that $\|\j_{it}(x)\|/\rho_t(x) <1$ for $i=1, \ldots, N$ \cite{bohm}, from which one easily deduces that $\B^\psi \ss \K_W$, and therefore the Bohmian space is a relativistic trajectory space. It is easy to see that the set $\B^\psi$ is covariant under the action of the subgroup $\PP_0 \ss \PP$ composed of the the translations and the 3-rotations, that is 
\begin{equation} \label{}
\B^{U_g\psi}= g \B^\psi \tx{ for all } g \in \PP_0.
\end{equation}
However the above equality does not hold in general when $g$ is a boost, and therefore the Bohmian space is not covariant \cite{bohm}. This determines the well-known tension existing between Bohmian mechanics and relativity theory.

Let us prove now the two results (a) and (b) announced in the introduction. The quantum system and the associated Bohmian space are assumed to be asymptotically regular (in the relativistic case it is assumed that the Bohmian space $(\B^{U_g\psi}, P^{U_g\psi})$ is asymptotically regular for all $g \in \PP$). As already said, for the quantum case the asymptotic regularity (the existence of the asymptotic velocity operators) has been proved for al large class of non-relativistic systems. The asymptotic regularity of the Bohmian space has been proved for a single non-relativistic particle under suitable conditions for the potential and the wave function \cite{asym}.

Under the assumption of asymptotical regularity the asymptotic measures $S^\psi_+$ (defined by equation (\ref{av})) and $Q^\psi_+$ (defined by equation (\ref{qap})) exist, and the first result to prove is that
\begin{equation} \label{fr}
S^\psi_+ = Q^\psi_+.
\end{equation}
In fact:
\bg{lista}
\item $S_+^\psi =\lim_{t \to \ity} S^\psi_t$ from the as. reg. of the Bohmian space (equation (\ref{prl1}));
\item $\lim_{t \to \ity} S^\psi_t=\lim_{t \to \ity} Q^\psi_t$ because $S^\psi_t = Q_t^\psi$ (equation (\ref{eq1})); 
\item $\lim_{t \to \ity} Q^\psi_t=Q^\psi_+$ from the as. reg. of the quantum system (equation (\ref{l4})).
\end{lista}
In the paper \cite{asym} mentioned above it has been proved that the distribution of the asymptotic velocities of the particle is given by the density $\|\hat \psi^\tx{out}(\cdot)\|^2 + \|\psi^{pp}\|^2 \de(\cdot)$, where $\hat \psi^\tx{out}$ is the Fourier transform of the outgoing asymptote of the scattering part of the wave function, and $\psi^{pp}$ is the bound part of the wave function. It is easy to see that this distribution corresponds to the distribution given by the asymptotic velocity operators; note for example that in this case $\hat \v_+=\hat \Om_+ \hat \v \hat \Om_+^\dg$, and $F_+(\{0\})$ is the projection onto the subspace corresponding to the pure point spectrum of the Hamiltonian (the bound states).

The second result to prove is that the asymptotic measure $S_+^\psi$ is covariant, that is
\begin{equation} \label{}
g S_+^\psi = S_+^{U_g \psi} \tx{ for all } g \in \PP,
\end{equation}
where $gS_+^\psi$ has been defined by equation (\ref{dav}). In fact:
\bg{lista}
\item $gS_+^\psi(V) = S^\psi_+(g^\mo V)$ from the as. reg. of the Bohmian space (equation (\ref{dav2}));
\item $S^\psi_+(g^\mo V)= Q^\psi_+(g^\mo V)$ from the first result (equation (\ref{fr})); 
\item $Q^\psi_+(g^\mo V)= Q^{U_g \psi}_+(V)$ from as. reg. of the quantum system (equation (\ref{mcv})); 
\item $Q^{U_g \psi}_+(V)=S^{U_g \psi}_+(V)$ from the first result.
\end{lista}

The relativistic non-covariance of the Bohmian space is usually expressed by saying that the construction of the Bohmian space requires a preferred foliation of spacetime. For example, the construction of the Bohmian space $(\B^\psi, P^\psi)$ is implicitly based on the foliation $\FF_0:=\{\{t\} \times \R^3\}_{t \in \R}$. On the other hand, the covariance of the asymptotic measure means that this measure does not depend on the foliation, as we can explicitly see with the following reasoning. The \Pe group acts naturally on the foliations, and any foliation can be expressed as $g \FF_0$, where $g$ is a suitable \Pe transformation. One can easily recognize that the Bohmian space generated by the wave function $\psi$ and by the foliation $g^\mo \FF_0$ is
\begin{equation} \label{}
g^\mo (\B^{U_g \psi}, P^{U_g \psi}).
\end{equation}
The asymptotic measure of this space is $g^\mo S_+^{U_g \psi}$ which, due to covariance, is equal to the asymptotic measure $S_+^\psi$ of the space $(\B^\psi, P^\psi)$ generated by the wave function $\psi$ and the foliation $\FF_0$.

\section{The covariant Bohmian space} \label{se5}

The covariance of the asymptotic probability allows us to build a covariant version of the Bohmian space as follows: (i) the set of the trajectories is the union of the sets of the Bohmian trajectories generated by all the (flat) foliations; (ii) the \sa is the one generated by the function $\eta_+$; (iii) it turns out that the value of probability measures associated with different foliations is the same on the elements of this $\si$-algebra. In order to simplifying the definition of the \sa $\si(\eta_+)$ (equation (\ref{asas}) below), in this section the set $\B^\psi$ will be redefined as $\B^\psi \sm \K_A$; this redefinition has no physical relevance because, under the assumption of asymptotic regularity, it consists of removing a set of zero measure form $\B^\psi$.

Let us introduce moreover the following notation:
\begin{equation} \label{}
\B^\psi_g := g^\mo \B^{U_g \psi} \tx{ and } P^\psi_g := g^\mo P^{U_g \psi};
\end{equation}
the Bohmian space $(\B^\psi_g, P^\psi_g)$ is therefore generated by the wave function $\psi$ and by the foliation $g^\mo \FF_0$. Let us define
\begin{equation} \label{}
\A^\psi:= \bigcup_{g \in \PP} \B^\psi_g.
\end{equation} 
One can easily prove that $\A^\psi$ is covariant, that is
\begin{equation} \label{}
g \A^\psi = \A^{U_g \psi}.
\end{equation}
In fact
\[
g \A^\psi = \bigcup_{h \in \PP} g h^\mo \B^{U_h \psi}= \bigcup_{f \in \PP} f^\mo \B^{U_{fg} \psi}= \bigcup_{f \in \PP} \B^{U_g \psi}_f = \A^{U_g \psi}.
\]
Let us define moreover the following probability measure on $(\A^\psi, \si(\A^\psi))$:
\begin{equation} \label{}
\bar P^\psi_g (K) := P^\psi_g (K \cap \B^\psi_g) \tx{ for all } K \in \si(\A^\psi) \tx{ and } g \in \PP.
\end{equation}
In words, $\bar P^\psi_g$ is the extension of $P^\psi_g$ from $\B^\psi_g$ to $\A^\psi$ defined by the condition $\bar P^\psi_g (\A^\psi \sm \B^\psi_g) =0$. The set $K \cap \B^\psi_g$ belongs to $\si(\B^\psi_g)$ because of equation (\ref{laste}). Let us introduce finally the $\si$-algebra
\begin{equation} \label{asas}
\AA^\psi:= \si(\eta_+|_{\A^\psi}).
\end{equation}
As shown in section \ref{se2} the map $\eta_+|_{\A^\psi}$ is measurable, and therefore $\AA^\psi \se \si(\A^\psi)$. Roughly speaking $\AA^\psi$ is a coarse-graining of $\si(\A^\psi)$ obtained from $\si(\A^\psi)$ by grouping the trajectories with the same asymptotic velocity. Again one can easily prove that $\AA^\psi$ is covariant:
\[
g \AA^\psi= \{g [\eta_+^\mo(V) \cap \A^\psi]: V \in \VV_1\} =\{\eta_+^\mo(g V) \cap \A^{U_g\psi}: V \in \VV_1\} = \AA^{U_g \psi}.
\]
Moreover one can prove that: (i) $\bar P^\psi_g$ and $ \bar P^\psi_{g'}$ have the same value on the elements of $\AA^\psi$, that is
\begin{equation} \label{}
\bar P^\psi_g(K)=\bar P^\psi_{g'}(K)=: \bar P^\psi(K) \tx{ for all } K \in \AA^\psi \tx{ and } g, g' \in \PP,
\end{equation}
and (ii) $\bar P^\psi$ is covariant, that is
\begin{equation} \label{}
g\bar P^\psi(K):= \bar P^\psi(g^\mo K) = \bar P^{U_g \psi}(K) \tx{ for all } K \in \AA^\psi \tx{ and } g \in \PP.
\end{equation}
In fact, if $K \in \AA^\psi$ there is a set $V \in \VV_1$ such that $K=\eta_+|_{\A^\psi}^\mo(V)$; so:
\bea
& & \bar P^\psi_g(K) = P^\psi_g(K \cap \B^\psi_g) = g^\mo P^{U_g \psi} (K \cap g^\mo \B^{U_g \psi}) =  \\
& & = P^{U_g \psi}[g \eta_+|_{\A^\psi}^\mo (V) \cap \B^{U_g \psi}] = P^{U_g \psi} \circ \eta_+|_{\B^{U_g \psi}}^\mo (g V) = S_+^{U_g \psi} (g V)= S_+^\psi(V),
\eea
which is independent of $g$ and covariant.

The covariant Bohmian space is therefore the probability space
\bg{equation} \label{cs}
 (\A^\psi, \AA^\psi, \bar P^\psi).
\end{equation}
Note that this space does not define an instantaneous probability distribution of the particles on configuration space; in fact the tentative definition 
\[
\bar P^\psi_t \stackrel{\tx{\tiny{tentative}}}{:=} \bar P^\psi \circ \pi_t|_{\A^\psi}^\mo,
\]
which is the analogous of the definition (\ref{isd}) of $P_t$, is not correct, because in general $\pi_t$ is not $\AA^\psi$-measurable.

The covariant space (\ref{cs}) is proposed here only as a mathematical entity. Its empirical adequacy is essentially based on the empirical adequacy of the \sa $\AA^\psi$, namely on the fact that two trajectories with the same asymptotic velocity are empirically indistinguishable. This will be discusses in a future paper. Another example of a covariant Bohmian space, based upon the notion of imprecise probability, has been proposed by the author in a previous paper \cite{galv}.

\appendix
\setcounter{secnumdepth}{0}

\section{Appendix} \nonumber

\bg{prop} \label{prop1}Let $\B \se \K$. If $\pi_t|_\B$ is bijective and $\pi_s \circ \pi_t|_\B^\mo:\X \to \X$ is measurable for all $s,t \in \R$, then $\pi_t|_\B^\mo$ is measurable for all $t \in \R$.
\end{prop}
\bg{proof} Since $\pi_t$ is bijective than $\pi_t|_\B^\mo$ is measurable between $\XX$, and $\si(\pi_t|_\B)$, so the thesis is proved if we prove that $\si(\B)=\si(\pi_t|_\B)$. Let $B \in \si(\pi_t|_\B)$ and let $X_t:=\pi_t(B)$; the set $X_s :=\pi_s(B)= \pi_s \circ \pi_t|_\B^\mo (X_t) \in \XX$ because the flux map is measurable, and therefore $B = \pi_s|_\B^\mo(X_s) \in \si(\pi_s|_\B)$. So $\si(\pi_t|_\B) \se \si(\pi_s|_\B)$. By symmetry one obtains the opposite relation, and therefore $\si(\B)=\si(\pi_t|_\B)$.
\en{proof}

\bg{prop} \label{prop2} Let $\A, \B \se \K$, with $\B \se \A$; then 
\bg{equation}
\si(\B)=\{\B \cap A: A \in \si(\A)\}.
\en{equation}
\en{prop}
\bg{proof} If $\FF$ is a generic class of subsets of $\A$ then $\si(\{ \B \cap A: A \in \FF\}) = \{\B \cap A : A \in \si(\FF)\}$ (this is valid for a generic set $\A$, see e.g. \cite{canna}, Exercise 1.12). So: $\si(\B) = \si(\{\B \cap \pi_t^\mo(X): X \in \XX \tx{ and } t \in \R\})=\si(\{\B \cap \A \cap \pi_t^\mo(X) : X \in \XX \tx{ and } t \in \R \})= \{\B \cap A: A \in \si(\A)\}$.
\en{proof}

\bg{thebibliography}{10}

\bibitem{bau} Baudoin, F.: Diffusion Processes and Stochastic Calculus. European Mathematical Society (2014)

\bibitem{billi} Billingsley, P.: Convergence of Probability Measures. Wiley, New York (1968)

\bibitem{bohm} Bohm, D., Hiley, B.J.: The Undivided Universe. Routledge, New York (1993)

\bibitem{canna} Cannarsa, P., D'Aprile, T.: Introduction to Measure Theory and Functional Analysis. Springer International Publishing, Switzerland (2015)

\bibitem{dere} Derezinski, J., Christian, G.: Scattering theory of classical and quantum N-particle systems. Springer Science \& Business Media (2013)

\bibitem{galv} Galvan, B.: Relativistic Bohmian mechanics without a preferred foliation. J. Stat. Phys. 161, 1268-1275 (2015) arxiv.org/abs/1509.03463

\bibitem{hyper} D\"urr, D., Goldstein, S., M\"unch-Berndl, K., Zangh\`i, N.: Hypersurface Bohm-Dirac models. Phys. Rev. A 60, 2729-2736 (1999) arXiv:quant-ph/9801070

\bibitem{dbook2} D\"urr, D., Goldstein, S., Zangh\`i, N.: Quantum Physics without Quantum Philosophy. Springer-Verlag, Berlin (2013)

\bibitem{qe} D\"urr, D., Goldstein, S., Zangh\`i, N.: Quantum equilibrium and the origin of absolute uncertainty. J. Statist. Phys. 67, 843–907 (1992) arXiv:quant-ph/0308039

\bibitem{fong} Fong, R., Sucher, J.: Relativistic Particles Dynamics and the $S$ Matrix. J. Math. Phys. 5, 456-470 (1964)

\bibitem{dbook} D\"urr, D., Teufel, S.: Bohmian Mechanics. Springer-Verlag, Berlin (2009)

\bibitem{hol} Holland, P.R.: The quantum Theory of Motion. Cambridge University Press, Cambridge
(1993)

\bibitem{reed} Reed, M., Simon, B.: Methods of Modern Mathematical Physics I: Functional Analysis. Academic press, San Diego (1980).

\bibitem{asym} R\"omer, S., D\"urr, D., Moser, T: Asymptotic behavior of Bohmian trajectories in scattering situations. J. Phys. A: Math. Gen. 38, 8421-8443 (2005) arXiv:math-ph/0505074

\bibitem{supp} Supplementary material: www.brunogalvan.it/sm01.pdf

\bibitem{global} Teufel, S., Tumulka, R.: Simple Proof for Global Existence of Bohmian Trajectories. Commun. Math. Phys. 258, 349-365 (2005) arXiv:math-ph/0406030

\bibitem{taller} Thaller, B.: The Dirac Equation. Springer-Verlag, Berlin (1992).

\end{thebibliography}

\end{document}